\documentclass[preprint,onecolumn,nofootinbib]{revtex4}
\pdfoutput=1
\usepackage[colorlinks=true,linkcolor=blue,urlcolor=blue,filecolor=black,citecolor=red,pdfstartview=FitV,pdftitle={},pdfsubject={},pdfkeywords={},pdfpagemode=None,bookmarksopen=true]{hyperref}
\usepackage{graphicx}
\usepackage{amsmath}
\usepackage{amsfonts}
\usepackage{amssymb,ulem}
\usepackage{color,xcolor}%
\usepackage{CJK}
\usepackage{subfigure}

\usepackage{dcolumn}
\newcommand{\f}{\begin{equation}}
\newcommand{\ff}{\end{equation}}
\newcommand{\fa}{\begin{eqnarray}}
\newcommand{\ffa}{\end{eqnarray}}

\begin{document}
\title{A Minkowski-core black hole with cosmological constant and electric charge}

\author{Wen-Zheng Chen$^{1}$}
\thanks{mx120240363@stu.yzu.edu.cn}
\author{Shulan Li$^{2}$}
\thanks{shulanli.yzu@gmail.com}
\author{Jian-Pin Wu$^{1}$}
\thanks{jianpinwu@yzu.edu.cn} 
\author{Cong Zhang$^{3}$}
\thanks{cong.zhang@bnu.edu.cn}
\affiliation{
 $^1$\mbox{Center for Gravitation and Cosmology, College of Physical Science and Technology,} \mbox{Yangzhou University, Yangzhou 225009, China}\\
$^{2}$\mbox{Department of Physics, Shanghai University, Shanghai 200444, China}
$^{3}$\mbox{School of Physics and Astronomy,} \mbox{Key Laboratory of Multiscale Spin Physics (Ministry of Education),}
\mbox{Beijing Normal University, Beijing 100875, China}}

\begin{abstract}
Within a covariant effective Hamiltonian framework, we employ an inverse construction to derive the gravitational Hamiltonian constraint for a Minkowski-core regular black hole without invoking exotic matter. We then extend the constraint by coupling it to a spherically reduced Maxwell field and including a cosmological constant. The resulting charged anti-de Sitter (AdS) and de Sitter (dS) solution is gauge independent and reduces to the Reissner--Nordstr\"om--AdS (RN-AdS) black hole when the regularization parameter vanishes. Electric charge and a cosmological constant preserve the Minkowski core: the metric approaches the Minkowski geometry at the center, the Kretschmann scalar vanishes there, and the spacetime exhibits a multi-horizon structure. Focusing on AdS backgrounds, we investigate the black hole thermodynamics. For a regularization parameter below a critical value, the model exhibits a phase transition with consistent signatures across these thermodynamic quantities, demonstrating that Minkowski-core regularization can preserve center regularity while modifying the AdS thermodynamic phase structure relative to the RN-AdS benchmark.
\end{abstract}

\maketitle
\tableofcontents

\section{Introduction}
Black holes are among the most striking predictions of general relativity (GR), and their existence has been confirmed by gravitational-wave observations \cite{LIGOScientific:2016aoc,LIGOScientific:2016lio,LIGOScientific:2025rid} and by the Event Horizon Telescope (EHT) \cite{EventHorizonTelescope:2019ths,EventHorizonTelescope:2019dse,EventHorizonTelescope:2019ggy}. The singularity theorems \cite{Hawking:1970zqf,Hawking2023TheLS,Senovilla:1998oua}, however, show that classical GR inevitably predicts essential singularities inside black holes, where geodesics become incomplete and curvature invariants diverge.

To circumvent spacetime singularities, numerous regular black hole models have been proposed \cite{2767662,Hayward:2005gi,Frolov:2016pav,Fan:2016hvf,Lan:2023cvz,Wang:2026sqr,Capozziello:2024ucm}. Following Bardeen's seminal construction \cite{2767662}, well-known examples include the Hayward \cite{Hayward:2005gi} and Frolov \cite{Frolov:2016pav} black holes, for which curvature invariants remain finite everywhere, thereby eliminating the central spacetime singularity. These geometries have been studied extensively, including their quasinormal modes \cite{Zhang:2025ygb,Song:2024kkx,Dong:2024ams,Zhang:2024nny}, thermodynamics \cite{Li:2018bny,Al-Badawi:2025uxp,Wu:2024gqi}, and particle trajectories \cite{Abdujabbarov:2016hnw,Zhang:2024hix,Zhang:2026lrd,Gong:2025mne,Battista:2026nsx}. However, the underlying dynamical theory responsible for these solutions was initially unknown. One natural approach is to attribute their regularity to the matter sector by considering classical GR coupled to suitable matter fields. For example, Ay\'on-Beato and Garc\'{\i}a demonstrated that the Bardeen black hole can be obtained as a solution of the Einstein equations coupled to nonlinear electrodynamics, where the source is interpreted as a magnetic monopole \cite{Ayon-Beato:2000mjt}. This provided the first well-defined action principle underlying the Bardeen solution. Similar constructions have subsequently been developed for many other regular black hole models \cite{Bronnikov:2017sgg,Nojiri:2017kex}. However, this strategy is widely viewed as unsatisfactory. Besides requiring exotic matter that typically violates the standard energy conditions, the underlying theories generally continue to admit singular black hole solutions in addition to the regular ones \cite{Huang:2025uhv,Rao:2025rop}.

An alternative approach is to attribute regularity to quantum-gravitational effects. Indeed, one of the central goals of quantum gravity is to resolve the spacetime singularities predicted by classical GR. Although a complete theory of quantum gravity has not yet been established, one can construct effective models incorporating quantum corrections that encode the low-energy consequences of the underlying quantum theory. Such effective descriptions have led to a variety of regular black hole solutions. Among the existing approaches, loop quantum gravity (LQG), a non-perturbative and background-independent quantization of GR \cite{Thiemann:2001gmi,Rovelli:1997yv,Ashtekar:2004eh,Han:2005km}, has achieved remarkable success in cosmology, giving rise to the well-developed framework of loop quantum cosmology \cite{Ashtekar:2011ni,Bojowald:2001xe}. Some ideas developed in loop quantum cosmology have also been extended to black holes, yielding a variety of singularity-free black hole models, collectively referred to as loop quantum black holes \cite{Ashtekar:2018lag,Alonso-Bardaji:2021yls,Zhang:2023yps,Perez:2017cmj,Lewandowski:2022zce}. These models are typically described by an effective Hamiltonian formulation rather than an effective Lagrangian. In constructing loop quantum black hole models, quantum gravity effects are commonly incorporated into the Hamiltonian constraint through holonomy corrections, which replace the connection variables by functions constructed from their holonomies, leading to modified effective dynamics. This replacement reflects the loop representation, ensuring background independence in the resulting quantum theory.
Various prescriptions for implementing holonomy corrections have been proposed (see, e.g., \cite{Modesto:2008im,Ashtekar:2018lag,Zhang:2023yps,Zhang:2021xoa,Husain:2022gwp,Chiou:2008eg}). However, not all such prescriptions preserve general covariance at the effective level \cite{Bojowald:2015zha}. Motivated by this issue, several generally covariant loop quantum black hole models have recently been developed \cite{Alonso-Bardaji:2022ear,Zhang:2024khj,Zhang:2024ney,Belfaqih:2024vfk}. Within these covariant effective theories, static spherically symmetric loop quantum black holes have been further generalized to include a cosmological constant and couplings to classical matter fields, such as electromagnetic and scalar fields \cite{Bojowald:2015zha,Alonso-Bardaji:2023niu,Zhang:2019acn}.

An important advantage of covariant loop quantum black hole models is that they resolve spacetime singularities without requiring exotic matter fields while preserving Birkhoff’s theorem, namely that the theory admits a unique static, spherically symmetric vacuum solution. They therefore avoid the main drawback of regular black holes constructed within classical GR coupled to matter. This naturally raises the following question: can one construct a covariant effective Hamiltonian description for regular black hole metrics that do not originate from loop quantum gravity? A recent work \cite{Zhang:2025ccx} provides an affirmative answer by proposing a general reconstruction procedure. Given a static, spherically symmetric metric, the method reconstructs an effective Hamiltonian constraint without introducing exotic matter, such that the prescribed metric is the unique static vacuum solution of the resulting theory. Moreover, the reconstructed theory is generally covariant. In the present work, we apply this method to the Minkowski-core black hole \cite{Ling:2021olm,Zeng:2023fqy,Xiang:2013sza} and derive its corresponding effective Hamiltonian. We then extend the model by coupling it to the Maxwell field and a cosmological constant, thereby obtaining a unified effective theory. Finally, we investigate the regularity and thermodynamic properties of the resulting black hole solutions.

The remainder of this paper is organized as follows. Sec.~\ref{section 2} reviews the covariant effective-Hamiltonian framework, constructs the Hamiltonian constraint for the Minkowski-core black hole, and obtains the charged (A)dS extension. In particular, its horizon structure, Kretschmann scalar, and regularity are analyzed in detail. Sec.~\ref{section 3} studies the thermodynamics of this model. Sec.~\ref{section 4} summarizes our results. Technical details of the reconstruction and the equations of motion are collected in Appendices~\ref{Appendix A} and~\ref{Appendix B}, and Appendix~\ref{Appendix C} presents the complete expression for the Kretschmann scalar, together with a note on the reality of the entropy. We set $G=c=\hbar=1$ throughout.

\section{Effective model}\label{section 2}

This section is divided into three subsections. Sec.~\ref{section 2A} introduces the covariant effective-Hamiltonian framework. Sec.~\ref{section 2B} constructs the Hamiltonian constraints for the Minkowski-core black hole, couples matter fields and analyzes the metric function. Sec.~\ref{section 2C} examines the Kretschmann scalar and the regularity of the extended spacetime.

\subsection{Framework of effective Hamiltonian} \label{section 2A}
\subsubsection{Structures of classical GR}
Let us start with the spherically symmetric sector of GR. In this case, the spacetime is based on a 4-dimensional manifold $\mathcal M\times \Sigma$ with $\Sigma\cong \mathbb \mathcal M_1\times\mathbb S^2$. Here $\mathcal M_1$ denotes the 1-dimensional manifold corresponding to the radial direction, and $\mathbb S^2$ is the 2-sphere. Let $(x,\theta,\phi)\in\mathcal M_1\times \mathbb S^2$ be the adapted coordinates.
In classical GR, the phase space of spherically symmetric gravity can be described in terms of the metric components $g_{xx}$ and $g_{\theta\theta}$, together with their time derivatives which determine the extrinsic curvature of the spatial hypersurfaces. In this paper, instead, we employ variables adapted to canonical loop quantum gravity, which are canonically equivalent to these metric variables. They are denoted by $(K_I,E^I)$ with $I=1,2$ \cite{Bojowald:1999eh,Bojowald:2005cb,Campiglia:2007pr}. The variables $E^I$ encode the intrinsic geometry of $\Sigma$, while $K_I$ encode its extrinsic curvature.
The non-vanishing Poisson brackets among the phase-space variables are
\begin{equation}
    \begin{aligned}[b]
		\{K_1(x), E^1(y)\} &= 2\delta(x,y), \\
		\{K_2(x), E^2(y)\} &= \delta(x,y).
        \label{eq1,poisson brackets}
	\end{aligned}
\end{equation}

In the canonical formulation, GR is a totally constrained system. This means that the dynamics of the model is encoded in a set of constraints, namely the Hamiltonian constraint $H^{\rm G}_{\rm cl}$ and the diffeomorphism constraint $H_x$. In the spherically symmetric sector, they take the form
\begin{equation}
	H_x= \frac{1}{2} (2E^2 \partial_x K_2 - K_1 \partial_x E^1),
    \label{eq2,diffeomorphism constraint}
\end{equation}
and 
\begin{equation}
    \begin{aligned}[b]
	H^{\rm G}_{\rm cl}= &-K_1 K_2 \sqrt{E^1} - \frac{E^2}{2 \sqrt{E^1}} - \frac{(K_2)^2 E^2}{2 \sqrt{E^1}} + \frac{(\partial_x E^1)^2}{8 \sqrt{E^1} E^2} \\
	&- \frac{\sqrt{E^1} (\partial_x E^1) \partial_x E^2}{2 (E^2)^2} + \frac{\sqrt{E^1} \partial_x^2 E^1}{2 E^2}.
     \end{aligned}
    \label{eq3,Hamiltonian constraint}
\end{equation}
The system is first class because the constraints form a closed algebra under the Poisson bracket. More explicitly, one finds
\begin{equation}
    \begin{aligned}[b]
        \left\{ H_{x}[N_{1}^{x}],H_{x}[N_{2}^{x}] \right\} &= H_{x}\left[ N_{1}^{x}\partial_{x}N_{2}^{x}-N_{2}^{x}\partial_{x}N_{1}^{x} \right], \\
	\left\{ H_{x}[N^{x}],H^{\rm G}_{\rm cl}[N] \right\} &= H^{\rm G}_{\rm cl}\left[ N^{x}\partial_{x}N \right], \\
	\left\{ H^{\rm G}_{\rm cl}[N_{1}],H^{\rm G}_{\rm cl}[N_{2}] \right\} &= H_{x}\left[ \frac{E^1}{(E^2)^2}\left( N_{1}\partial_{x}N_{2}-N_{2}\partial_{x}N_{1} \right) \right],
    \end{aligned}
    \label{eq4,constraint algebracl}
\end{equation}
where we used the convention $G[f]\equiv \int {\rm d}x f(x)G(x)$. To solve the dynamics, one needs to choose a lapse function $N$ and a shift vector $N^x$. The time evolution is then generated through the Hamilton equations
\begin{equation}
\begin{aligned}[b]
\partial_t E^I &= \left\{E^I, H^{\rm G}_{\rm cl}[N]+H_x[N^x]\right\},\\
\partial_t K_I &= \left\{K_I, H^{\rm G}_{\rm cl}[N]+H_x[N^x]\right\}.
\label{eq5,time evolution}
\end{aligned}
\end{equation}
These equations must be solved with initial data satisfying the constraints $H^{\rm G}_{\rm cl}=0=H_x$. The corresponding spacetime metric is given by
\begin{equation}
g_{\mu\nu} {\rm d}x^\mu {\rm d}x^\nu = -N^2 {\rm d}t^2 + \frac{(E^2)^2}{ E^1} ({\rm d}x + N^x {\rm d}t)^2 + E^1({\rm d}\theta^2 + \sin^2 \theta {\rm d}\phi^2 ). 
\label{eq6,classical metric}
\end{equation}
Two remarks are in order. First, because the constraint algebra is closed, the constraints are preserved by the time evolution: if they hold on the initial slice, they remain satisfied on any later time slice. Second, the resulting spacetime metric is invariant up to spacetime diffeomorphisms, reflecting the covariance of the theory.
 
\subsubsection{Extensions to effective model}\label{sec:solvdynamics}
We now turn to the effective model introduced in \cite{Zhang:2024khj,Zhang:2024ney}. This model can be regarded as a modified theory of gravity formulated in the Hamiltonian framework. It is effective in the sense that the Hamiltonian constraint is modified from its classical expression, referred to as $H^{\rm G}_{\rm eff}$, while the phase-space structure, as well as the diffeomorphism constraint, remains the same as in the classical spherically symmetric theory. Since the diffeomorphism constraint is kept in its classical form, the Poisson brackets
$\left\{H_x[N_1^x],H_x[N_2^x]\right\}$ and $\left\{H_x[N^x],H^{\rm G}_{\rm eff}[N]\right\}$
retain the same form as the first two relations in $\eqref{eq4,constraint algebracl}$, with the classical Hamiltonian constraint $H^{\rm G}_{\rm cl}$ replaced by the effective Hamiltonian constraint $H^{\rm G}_{\rm eff}$. Therefore, to ensure the closure of the constraint algebra, the only remaining bracket to be specified is
$\left\{ H^{\rm G}_{\rm eff}[N_{1}],H^{\rm G}_{\rm eff}[N_{2}] \right\}$, which is assumed to be 
\begin{equation}
    \begin{aligned}[b]
	\left\{ H^{\rm G}_{\rm eff}[N_{1}],H^{\rm G}_{\rm eff}[N_{2}] \right\} &= H_{x}\left[ \frac{\mu E^1}{(E^2)^2}\left( N_{1}\partial_{x}N_{2}-N_{2}\partial_{x}N_{1} \right) \right].
    \end{aligned}
    \label{eq7,constraint algebra}
\end{equation}
Here, the factor $\mu$ is introduced as a general phase-space function due to the modification of the Hamiltonian constraint. Because of the presence of this factor, in order for the constraint algebra to retain the geometric interpretation of hypersurface deformations, the spacetime metric is defined as
\begin{equation}
	g_{\mu\nu}^{\rm eff} {\rm d}x^\mu {\rm d}x^\nu = -N^2 {\rm d}t^2 + \frac{(E^2)^2}{\mu E^1} ({\rm d}x + N^x {\rm d}t)^2 + E^1({\rm d}\theta^2 + \sin^2 \theta {\rm d}\phi^2 ).
    \label{eq8,general metric}
\end{equation}

In the Hamiltonian formulation, one has to perform a $3+1$ decomposition of spacetime, so covariance is no longer manifest. Consequently, the requirement of general covariance imposes nontrivial restrictions on the form of $H^{\rm G}_{\rm eff}$. In \cite{Zhang:2024khj,Zhang:2024ney}, the issue of general covariance in the Hamiltonian formulation was investigated in details. It turns out that, to have covariance, i.e, the effective metric \eqref{eq8,general metric} is invariant up to diffeomorphism transformations for various choice of lapse function $N$ and shift vector $N^x$, $H^{\rm G}_{\rm eff}$ needs to takes the form: 
\begin{equation}
		H^{\rm G}_{\rm eff} = -2E^2 \left[ \partial_{s_1} M_{\rm eff} + \frac{\partial_{s_2} M_{\rm eff}}{2} s_3 + \frac{\partial_{s_4} M_{\rm eff}}{s_4} s_5 + \mathcal{R} \right],
        \label{eq9,covarianc Hamiltonian}
\end{equation}
where $$s_1 = E^1,\quad s_2 = K_2,\quad s_3 = \frac{K_1}{E^2},\quad s_4 = \frac{\partial_x s_1}{E^2},\quad s_5 = \frac{\partial_x s_4}{E^2},$$ are phase space dependent variables, 
and $\mathcal{R}$ is an arbitrary function of $s_1$ and $M_{\rm eff}$. $M_{\rm eff}$ is referred to as the effective mass function, which depends on $s_1, s_2, s_4$. $M_{\rm eff}$ and $\mu$ must satisfy the following system of equations, called the covariance equations.
\begin{equation}
	\frac{\mu s_1 s_4}{4} = (\partial_{s_2} M_{\rm eff}) \partial_{s_2} \partial_{s_4} M_{\rm eff} - (\partial_{s_4} M_{\rm eff}) \partial_{s_2}^2 M_{\rm eff},
    \label{eq10,covariance equations 1}
\end{equation}
and
\begin{equation}
	0 = (\partial_{s_2} \mu) \partial_{s_4} M_{\rm eff} - (\partial_{s_2} M_{\rm eff}) \partial_{s_4} \mu.
    \label{eq11,covariance equations 2}
\end{equation}
Once $M_{\rm eff}$ and $\mu$ solving Eqs.~\eqref{eq10,covariance equations 1} and \eqref{eq11,covariance equations 2} are specified, Eq.~\eqref{eq9,covarianc Hamiltonian} determines $H^{\rm G}_{\rm eff}$ up to the choice of $\mathcal{R}$. Setting $\mathcal{R}=0$ makes $M_{\rm eff}$ a Dirac observable interpretable as the black-hole mass parameter. The free choice of $\mathcal R$, together with the fact that solutions to \eqref{eq10,covariance equations 1} and \eqref{eq11,covariance equations 2} are not unique, implies that that covariance does not fix a unique $H^{\rm G}_{\rm eff}$.

Once $H^{\rm G}_{\rm eff}$ is obtained, the dynamics can in principle be solved as in classical GR: one chooses a lapse function $N$ and a shift vector $N^x$, and then solves the Hamilton equations. In practice, however, we usually adopt a reverse strategy: instead of choosing $N$ and $N^x$ from the outset, we first impose two appropriate gauge-fixing conditions. These conditions may be specified by prescribing $E^I$, $K_I$, or suitable combinations of the phase-space variables. As an example, let us consider the gauge fixing in which $E^1(t,x)$ and $E^2(t,x)$ are prescribed. Then, using the Hamilton equations
$\partial_t E^I=\{E^I,H^{\rm G}_{\rm eff}[N]+H_x[N^x]\}$,
we determine the corresponding lapse function $N$ and shift vector $N^x$. The remaining phase-space variables $K_I(t,x)$ are then obtained by substituting the prescribed functions $E^I(t,x)$ into the constraints. 
Different choices of $E^I$ correspond to different coordinate systems. For example, choosing $E^1=x^2$ and $E^2=x$ leads to the metric in Painlev\'e--Gullstrand (PG) coordinates. On the other hand, choosing $E^1=x^2$ together with $\partial_{s_2}M_{\rm eff}=0$ yields the metric in Schwarzschild coordinates. 
Following this strategy, Ref. \cite{Zhang:2025ccx} solved the vacuum dynamics for the general Hamiltonian constraint \eqref{eq9,covarianc Hamiltonian} and obtained the corresponding static spherically symmetric metric. More importantly, it showed that the procedure can be reversed: starting from an arbitrary static spherically symmetric metric, one can reconstruct a generally covariant effective Hamiltonian constraint for which the prescribed metric is the unique static vacuum solution. Since this reconstruction procedure forms the basis of the present work, we briefly review it in Appendix \ref{Appendix A}, where the construction of the corresponding $M_{\rm eff}$ and Hamiltonian constraint is summarized.

\subsection{Hamiltonian constraint and matter coupling for the Minkowski-core black hole} \label{section 2B}

In this subsection, we constructed the Hamiltonian constraint for the Minkowski-core black hole and couple matter fields to obtain the extended model. Frist, we derive the effective Hamiltonian constraint of Minkowski-core black hole. The metric in Schwarzschild coordinates is expressed as
\begin{equation}
    \begin{aligned}[b]
        \mathrm{d}s^2 &= -f_0(x)\mathrm{d}t_s^2 +\frac{1}{f_0(x)}\mathrm{d}x^2 +x^2\mathrm{d}\Omega^2; \\
    f_0(x)&= 1-\frac{2 M e^{-\frac{\alpha M^z}{x^n}}}{x},
    \end{aligned}
    \label{eq12,mc function}
\end{equation}
with $n\geq1$, $n>z \geq 0$ and $\alpha>0$, where $\alpha$ is the regularization parameter \cite{Ling:2021olm}. In this study, we only consider the case $z=0$, since the effective Hamiltonian constraint cannot be solved explicitly when $z \neq 0$. Following the method proposed in \cite{Zhang:2025ccx}, we obtain the effective mass $M_{\rm eff}$ and Hamiltonian constraint $H^{\rm G}_{\rm eff}$ of this model.
\begin{equation}
    	M_{\rm eff}=\frac{1}{8} \sqrt{{s_1}} e^{\alpha  \left(-{s_1}^{-n/2}\right)} \left[4 {s_2}^2-\left({s_4}^2-4\right) e^{2 \alpha  {s_1}^{-n/2}}\right],
        \label{eq13,mc Meff}
\end{equation}
\begin{equation}
    \begin{aligned}[b]
H^{\rm G}_{\rm eff} 
=& \frac{(E^1)^{-n/2-1} e^{-\alpha  (E^1)^{-n/2}}}{8 (E^2)^2}
   \Big\{ (E^1)^{(n+1)/2} \Big[E^2(\partial_x E^1)^2 e^{2 \alpha  (E^1)^{-n/2}}  \\
& -4 (E^2)^3 \left(e^{2 \alpha  (E^1)^{-n/2}}+(K_2)^2\right) \Big]  + \alpha  n \sqrt{E^1} (E^2) \Big[ 4 (E^2)^2 \left(e^{2 \alpha  (E^1)^{-n/2}}-(K_2)^2\right)  \\
& -(\partial_x E^1)^2 e^{2 \alpha  (E^1)^{-n/2}} \Big]  -4 (E^1)^{\frac{n+3}{2}} \Big[ -E^2 (\partial_x^2 E^1) e^{2 \alpha  (E^1)^{-n/2}}  \\
& +(\partial_x E^1) (\partial_x E^2) e^{2 \alpha  (E^1)^{-n/2}}   +2 (E^2)^2 K_1 K_2 \Big] \Big\}.
    \end{aligned}
    \label{eq14,mc Heff}
\end{equation}
We have constructed the corresponding Hamiltonian constraint for this regular black hole from the gravitational canonical variables alone. Setting $\alpha=0$ reduces Eq.~\eqref{eq14,mc Heff} to the classical Schwarzschild constraint \eqref{eq3,Hamiltonian constraint}. The equations of motion are given in Appendix~\ref{Appendix B}. For different gauge choices, the resulting line element can always be transformed into the form \eqref{eq12,mc function}.

Coupling matter fields to the gravitational sector, the total Hamiltonian constraint reads
\begin{equation}
	H^{\rm T}_{\rm eff}=H^{\rm G}_{\rm eff}+H^{\rm M}_{\rm eff},
    \label{eq15,total Hamiltonian constraint }
\end{equation}
where $H^{\rm G}_{\rm eff}$ and $H^{\rm M}_{\rm eff}$ denote the gravitational and matter contributions, respectively, and $H_x^{\rm T}$ is the total diffeomorphism constraint. The total Hamiltonian is given by $\mathbb{H}=\int dx\,(N H^{\rm T}_{\rm eff}+N^x H_x^{\rm T})$. We first include a cosmological constant $\Lambda$, which contributes only to the matter Hamiltonian constraint,
\begin{equation}
	H^{\rm \Lambda}_{\rm eff}=\frac{1}{2} \sqrt{E^1} E^2 \Lambda,
    \label{eq16,cosmological constant}
\end{equation}
while its diffeomorphism constraint vanishes \cite{Alonso-Bardaji:2023niu}.

For the Maxwell field \cite{Tibrewala:2012xb,Yang:2025ufs}, we introduce an additional pair of canonical variables: the electromagnetic four-potential $A_{\mu}$ and its conjugate momentum $p^{\mu}$. Spherical symmetry, together with the primary constraint $p^0=0$, reduces the electromagnetic phase space to $\{A_1,p^1\}$, with Poisson bracket
\begin{equation}
	\{A_1(x), p^1(y)\} = \delta(x,y).
    \label{eq17,em poisoon barcket}
\end{equation}
The electromagnetic sector is subject to three constraints: the Hamiltonian constraint $H^{\rm EM}$, the diffeomorphism constraint $H^{\rm EM}_x$, and the Gauss constraint $\mathcal{G}$:
\begin{equation}
	\begin{aligned}[b]
	  H^{\rm EM}&=\frac{E^2 (p^1)^2}{2(E^1)^{3/2}} + \frac{\sqrt{E^1}}{E^2} A_1 \mathcal{G},  \\
	H_x^{\rm EM}&= -A_1 \partial_x p^1,\\
	\mathcal{G}&=-\partial_x p^1.
	\end{aligned}
    \label{eq18,em constraints}
\end{equation}
The Gauss constraint $\mathcal{G}=0$ means that $p^1$ is a constant, which defines the electric charge $Q$. With this choice, the electromagnetic Hamiltonian constraint reduces to 
\begin{equation}
	H^{\rm EM} \big|_{p^{1}=Q} = \frac{E^2 Q^2}{2(E^1)^{3/2}}.
    \label{eq19,em constrain}
\end{equation}
Substituting the Hamiltonian constraints from \eqref{eq19,em constrain} and \eqref{eq16,cosmological constant} into \eqref{eq14,mc Heff}, we obtain the total Hamiltonian constraint for the black hole model coupled with matter fields
\begin{equation}
     H^{\rm T}_{\rm eff}=H^{\rm G}_{\rm eff}+\frac{1}{2} \sqrt{E^1} E^2 \Lambda+\frac{E^2 Q^2}{2(E^1)^{3/2}},
     \label{eq20,model total Heff}
\end{equation}
where $H^{\rm T}_{\rm eff}$ denotes the full Hamiltonian constraint including charge and $\Lambda$.

Now, we turn to work out the charged (A)dS extension of the Minkowski-core black hole by solving the equations of motion generated by \eqref{eq20,model total Heff}; the detailed derivation is given in Appendix~\ref{Appendix B}. In the Schwarzschild gauge ($E^1=x^2$, $\partial_{s_2}M_{\rm eff}=0$), the equations of motion are solved in Appendix~\ref{Appendix B}; substituting the resulting $E^2$ and $N$ (with $N^x=0$) into the general metric \eqref{eq8,general metric} yields the line element
\begin{equation}
    \begin{aligned}[b]
    \mathrm{d}s_{\rm eff}^2 &= -f(x)\mathrm{d}t_s^2 +\frac{1}{f(x)}\mathrm{d}x^2 +x^2\mathrm{d}\Omega^2,\\
    	f(x)&=1-\frac{2 m e^{\alpha  \left(-x^{-n}\right)}}{x}+\frac{Q^2 e^{\alpha  \left(-x^{-n}\right)}}{x^2}-\frac{1}{3} \Lambda  x^2 e^{\alpha  \left(-x^{-n}\right)}. 
    \label{eq21,new metric}
    \end{aligned}
\end{equation}
In the limit $\alpha\to 0$, Eq.~\eqref{eq21,new metric} reduces to the RN-(A)dS metric function
\begin{equation}
    f_{\text{RN-(A)dS}}(x)=1-\frac{2m}{x}+\frac{Q^2}{x^2}-\frac{\Lambda x^2}{3},
    \label{eq22,RNAdS metric}
\end{equation}
which serves as the benchmark for the thermodynamic analysis in Sec.~\ref{section 3}.

Solving the same equations in the PG gauge ($E^1=x^2$, $E^2=x$)---see Appendix~\ref{Appendix B} for details---and inserting the PG solutions into \eqref{eq8,general metric} yields, in coordinates $(t_p,x,\theta,\phi)$,
\begin{equation}
    \begin{aligned}[b]
	{\rm d}s_{\rm eff}^{2}&=-(1-\left(N^{x}_{p}\right)^{2}){\rm d}t_{p}^{2}+2N^{x}_{p}{\rm d}x{\rm d}t_{p}+{\rm d}x^{2}+x^{2}{\rm d}\Omega ^{2},\\
    N^{x}_{p}&=-\frac{e^{-\frac{1}{2} \alpha  x^{-n}} \sqrt{2 m x-Q^2+\frac{\Lambda  x^4}{3}}}{x},
    \label{eq23,new metric PG}
    \end{aligned}
\end{equation}
where $N^x_p$ is the shift vector in PG coordinates. Since the effective Hamiltonian constraint \eqref{eq20,model total Heff} is covariant, the forms of the line element in the two gauges differ only by a coordinate transformation
\begin{equation}
    {\rm d}{t_s} = {\rm d}{t_p} - \frac{N^{x}_{p}}{1 - (N^{x}_{p})^{2}} {\rm d}x.
    \label{eq24,coordinate trans}
\end{equation}
Therefore, the line elements in Eqs.~\eqref{eq21,new metric} and \eqref{eq23,new metric PG} describe the same geometry, confirming that the extended metric \eqref{eq21,new metric} is gauge-independent.

\begin{figure}[h]
    \centering
    \includegraphics[width=0.49\textwidth]{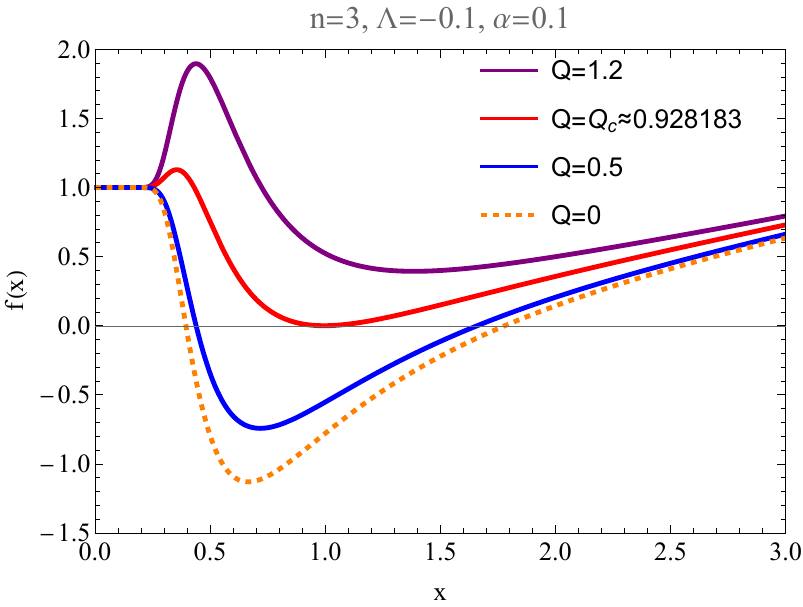}
    \hfill
    \includegraphics[width=0.49\textwidth]{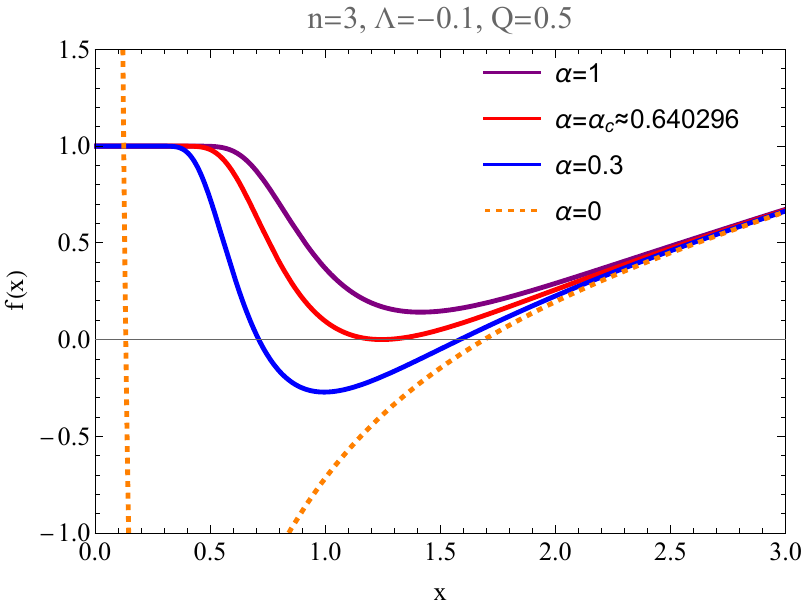}
    \hfill
    \caption{Plots of the metric function $f(x)$ versus $x$ with $n=3$. For the left panel, we choose $\Lambda=-0.1,\alpha=0.1$, and $Q=0,0.5,0.928183,1.2$, where $Q=Q_c\approx 0.928183$ is the extremal value with a single degenerate horizon; for the right panel, we choose $\Lambda=-0.1,Q=0.5$, and $\alpha= 0,0.3,0.640296,1$, where $\alpha=\alpha_c\approx0.640296$ is the corresponding extremal value.}
    \label{P1:model metric}
\end{figure}

We now analyze the horizon structure of the geometry derived above. Without loss of generality, we discuss the plot of the metric function by taking $n=3$. Fig.~\ref{P1:model metric} shows $f(x)$ for negative $\Lambda$, with $m=1$ throughout. When $\alpha \neq 0$, the metric function approaches unity at small $x$, which is the defining feature of the Minkowski-core black hole; neither $Q$ nor $\Lambda$ spoils this behavior near $x=0$, so no horizon appears in that region. At large radius the profile resembles that of the RN-AdS black hole. Event horizons correspond to positive roots of $f(x)=0$. With $\alpha$ fixed, increasing $Q$ drives a standard extremal transition: for $Q<Q_c$ there are two distinct horizons; at $Q=Q_c$ the inner and outer horizons merge into a single degenerate root; and for $Q>Q_c$ no horizon exists. For the parameters in the left panel, $Q_c\approx0.928183$. Likewise, with $Q$ fixed, increasing $\alpha$ produces the same $2\to1\to0$ pattern, with $\alpha_c\approx0.640296$ in the right panel. When $\alpha=0$, the solution reduces to Eq.~\eqref{eq22,RNAdS metric}, for which $f(x)$ diverges at $x=0$.

\begin{figure}[h]
    \centering
    \includegraphics[width=0.49\textwidth]{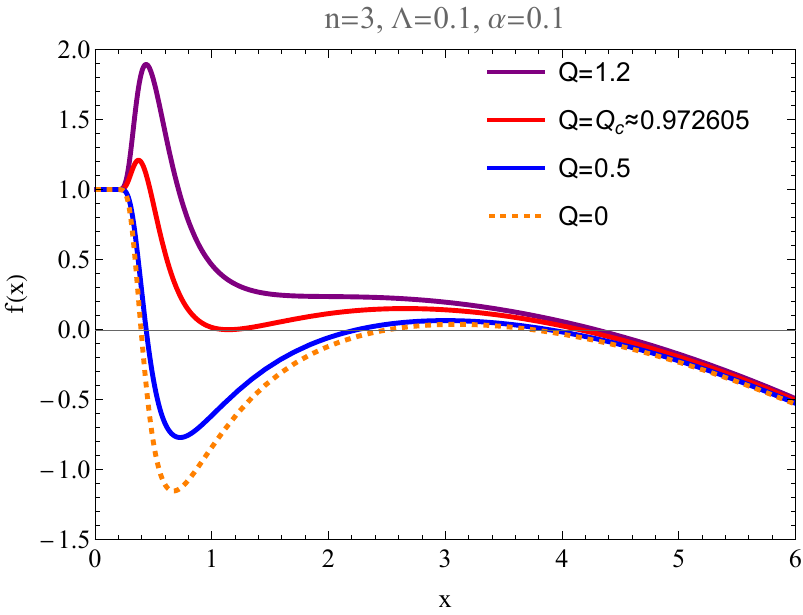}
    \hfill
    \includegraphics[width=0.49\textwidth]{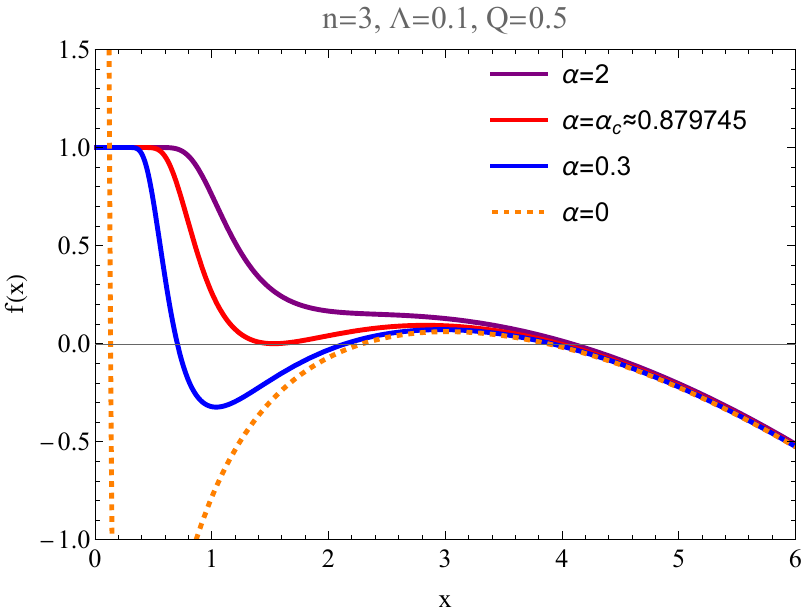}
    \hfill
    \caption{Plots of the metric function $f(x)$ versus $x$ with $n=3$. For the left panel, we choose $\Lambda=0.1,\alpha=0.1$, and $Q=0,0.5,0.972605,1.2$, where $Q=Q_c\approx 0.972605$ is the extremal value with a single degenerate horizon; for the right panel, we choose $\Lambda=0.1,Q=0.5$, and $\alpha= 0,0.3,0.879745,2$, where $\alpha=\alpha_c\approx0.879745$ is the corresponding extremal value.}
    \label{P2:model metric positive}
\end{figure}
Fig.~\ref{P2:model metric positive} shows the case of a positive cosmological constant. The left panel corresponds to a fixed parameter $\alpha$, and the right panel to a fixed charge parameter $Q$. One can observe that for small $x$, the behavior of the metric function is similar to that in the case of a negative cosmological constant, while for large $x$, a new root appears in the metric function, which is generally identified as the cosmological horizon.

\subsection{Curvature invariants and regularity of the extended model}\label{section 2C}

Having constructed the extended metric \eqref{eq21,new metric} and verified its gauge independence, we now assess the geometric regularity of the spacetime through the Kretschmann scalar. For a spherically symmetric line element, curvature singularities manifest as divergences of $\mathcal{K}=R_{\mu\nu\rho\sigma}R^{\mu\nu\rho\sigma}$, whereas a finite or vanishing $\mathcal{K}$ at small radius signals the absence of a classical curvature singularity. This analysis complements the horizon study of Sec.~\ref{section 2B} and establishes whether the introduction of $Q$ and $\Lambda$ preserves the Minkowski-core regularity of the vacuum model.

The Kretschmann scalar $\mathcal{K}$ of the extended model is
\begin{equation}
    \begin{aligned}[b]
\mathcal{K}&=\frac{1}{9} x^{-4 (n+2)} e^{-2 \alpha  x^{-n}} \mathcal{P}(x,x^n),
    \end{aligned}
    \label{eq25,model K scalar}
\end{equation}
where $\mathcal{P}$ is a polynomial in $x$ and $x^n$, with the general term structure $\mathcal{P}(x,x^n) = \sum_{i,j} C_{i,j}\, x^{i + j n}$. The coefficients $C_{i,j}$ depend on the model parameters $m$, $Q$, $\Lambda$, $\alpha$, and $n$, and all terms satisfy $i+jn \ge 0$. Under the assumptions $\alpha, n>0$, the exponential factor dominates as $x\to0$, so that $\mathcal{K}$ tends to zero exponentially fast. In the limit $x\to\infty$, $e^{-2\alpha x^{-n}}\to1$. Among the terms in $\mathcal{P}$, only the $x^{8+4n}$ term (whose coefficient is proportional to $\Lambda^2$) combines with the prefactor to produce a constant limit; all other contributions are of negative power and vanish. The full expression for the Kretschmann scalar is given in Appendix \ref{Appendix C}. Consequently, for generic non-vanishing parameter values, the leading behavior at the origin and at spatial infinity is
\begin{equation}
	\lim_{x \to 0} \mathcal{K} = 0;
	\lim_{x \to \infty} \mathcal{K} = \frac{8 \Lambda ^2}{3}.
    \label{eq26,K scalar limit}
\end{equation}
These limits imply that $\mathcal{K}\to0$ as $x\to0$ for generic parameters, so the introduction of $Q$ and $\Lambda$ does not spoil the Minkowski core, while at spatial infinity, $\mathcal{K}$ approaches the finite value set by $\Lambda$.

\begin{figure}[h]
    \centering
    \includegraphics[width=0.32\textwidth]{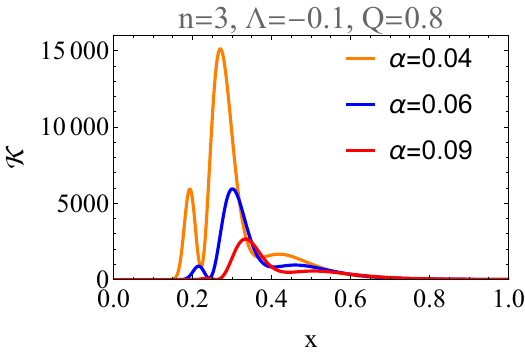}
    \hfill
    \includegraphics[width=0.32\textwidth]{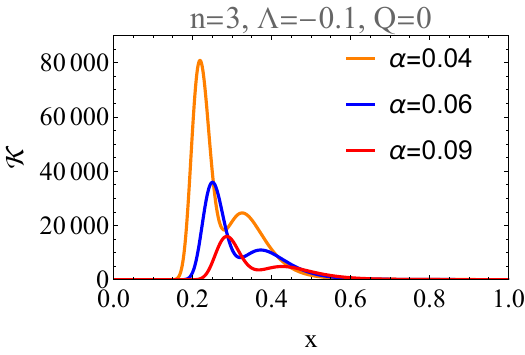}
    \hfill
    \includegraphics[width=0.32\textwidth]{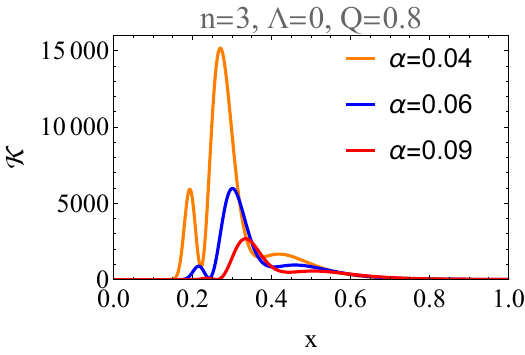}
    \caption{{Plots of the Kretschmann scalar $\mathcal{K}$ versus $x$ of the model with $n=3$: the left plot uses parameter $\Lambda=-0.1,Q=0.8$ and gradually increases the value of $\alpha=0.04,0.06,0.09$; the middle plot uses parameters $\Lambda=-0.1,Q=0$ and $\alpha=0.04,0.06,0.09$; the right plot uses parameters $\Lambda=0,Q=0.8$ and $\alpha=0.04,0.06,0.09$.}}
    \label{P3,model K scalar}
\end{figure}

Fig.~\ref{P3,model K scalar} illustrates these asymptotic features for $n=3$ and displays the intermediate-$x$ behavior of $\mathcal{K}$. In the left panel both $Q$ and $\Lambda$ are nonzero, in the middle panel $Q=0$, and in the right panel $\Lambda=0$. In all cases, the curves confirm that $\mathcal{K}$ vanishes at the origin and approaches the constant in \eqref{eq26,K scalar limit} at large $x$, while developing multiple peaks at intermediate radii.

\section{Black hole thermodynamics}\label{section 3}
In this section we analyze the thermodynamics \cite{Bardeen:1973gs,Bekenstein:1973ur,Wald:1979zz} of the extended model, focusing on the entropy, Hawking temperature, heat capacity and free energy  \cite{Lan:2023cvz,Zhang:2016ilt,Ahmed:2026bwm}. We denote the horizon radius by $x_h$ and focus on negative $\Lambda$, where the thermodynamic phase structure is richest. The cosmological constant is related to the thermodynamic pressure through $P=-\Lambda/8\pi$. With the additional regularization parameter $\alpha$, the first law takes the form
\begin{equation}
	{\rm d}M=T{\rm d}S+V{\rm d}P+\Phi_{Q}{\rm d}Q+\Phi_{\alpha}{\rm d}\alpha.
    \label{eq27,frist law}
\end{equation}
In the present analysis we treat $Q$, $\Lambda$, and $\alpha$ as fixed, so that ${\rm d}Q={\rm d}P={\rm d}\alpha=0$ and the first law reduces to ${\rm d}M=T{\rm d}S$. The entropy is obtained from the mass-temperature relation through
\begin{equation}
    S = \int \frac{{\rm d}M}{T}.
    \label{eq28,entropy}
\end{equation}

For the model \eqref{eq21,new metric}, the condition $f(x_h)=0$ determines the black-hole mass as a function of the horizon radius,
\begin{equation}
    	M=\frac{3 x_h^2 e^{\alpha  x_h^{-n}}-\Lambda  x_h^4+3 Q^2}{6 x_h}.
        \label{eq29,model mass}
\end{equation}
The conjugate potentials associated with Eq.~\eqref{eq27,frist law} are
\begin{align}
    \Phi_Q &= \left(\frac{\partial M}{\partial Q}\right)_{x_h,\Lambda,\alpha}=\frac{Q}{x_h}, \label{eq30,Phi Q}\\
    \Phi_\alpha &= \left(\frac{\partial M}{\partial \alpha}\right)_{x_h,Q,\Lambda}=\frac{1}{2} x_h^{1-n} e^{\alpha  x_h^{-n}}, \label{eq31,Phi alpha}\\
    V &= -8\pi\left(\frac{\partial M}{\partial \Lambda}\right)_{x_h,Q,\alpha}=\frac{4\pi x_h^3}{3}. \label{eq32 V}
\end{align}
These relations reduce to the standard RN-AdS expressions when $\alpha=0$.
Black hole radiation in a static spacetime is described by the Hawking temperature, which is related to the surface gravity $\kappa$ at the black hole horizon
\begin{equation}
	T = \frac{\mathcal{\kappa}}{2\pi} = \left. \frac{f'(x)}{4\pi}  \right|_{x=x_h},
    \label{eq33,temperature}
\end{equation}
where a prime denotes differentiation with respect to the radial coordinate. Explicitly,
\begin{equation}
    	T = -\frac{Q^2 e^{-\alpha  x_h^{-n}}}{4 \pi  x_h^3}-\frac{\Lambda  x_h e^{-\alpha  x_h^{-n}}}{4 \pi }-\frac{\alpha  n x_h^{-n-1}}{4 \pi }+\frac{1}{4 \pi  x_h}.
        \label{eq34,model hawking T}
\end{equation}
\begin{figure}[h]
    \centering
    \hfill
    \includegraphics[width=0.49\textwidth]{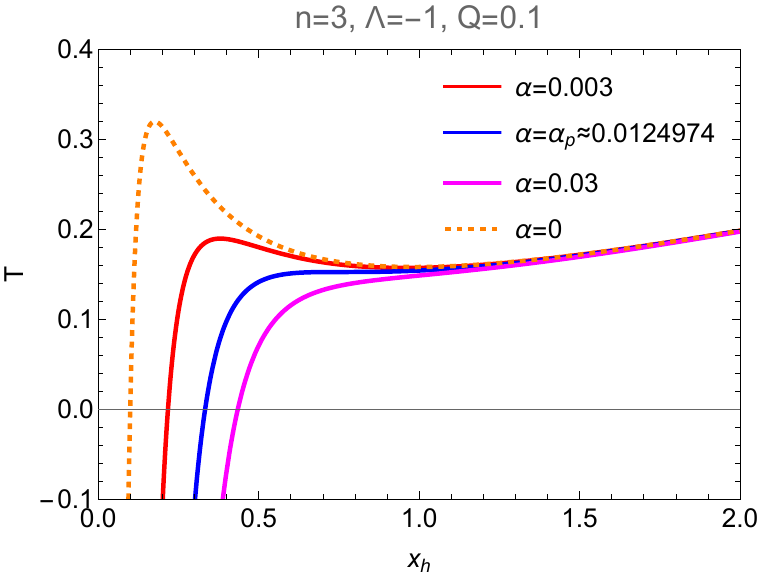}
    \hfill
    \includegraphics[width=0.49\linewidth]{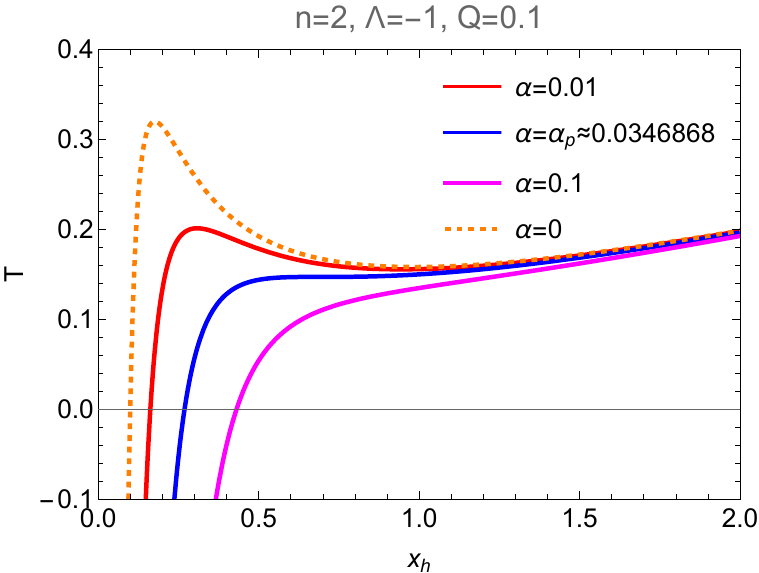}
    \caption{Plots of the Hawking temperature $T$ of the model as a function of $x_h$. In the left plot, the parameters are fixed at $n=3, \Lambda=-1, Q=0.1$, while $\alpha=0,0.003,0.0124974,0.03$. In the right plot, the parameters are fixed at $n=2, \Lambda=-1, Q=0.1$, while $\alpha=0,0.01,0.0346868,0.1$.}
    \label{P4,model hawking T}
\end{figure}

In Fig.~\ref{P4,model hawking T}, we take $n=3$ and $n=2$ as examples to illustrate the behavior of the Hawking temperature $T$ as a function of $x_h$ at fixed $Q$ and $\Lambda$. For sufficiently small $\alpha$ (red curves), $T$ first rises sharply to a maximum, then decreases slowly to a minimum, and finally increases monotonically with $x_h$ again, closely resembling that of the RN-AdS black hole (orange dashed curves). Obviously, $T(x_h)$ features two turning points: a maximum followed by a minimum. As the parameter $\alpha$ gradually increases and exceeds a critical value $\alpha_p$, the Hawking temperature increases monotonically with $x_h$ without any turning points (magenta curves). 

The critical value $\alpha_p$ is determined by
\begin{equation}
    \frac{\partial T}{\partial x_h} = 0, \quad \frac{\partial^2 T}{\partial^2 x_h} = 0.
    \label{eq35,turning point}
\end{equation}
For the parameters chosen in Fig.~\ref{P4,model hawking T}, $\alpha_p\approx0.0124974$ when $n=3$ and $\alpha_p\approx0.0346868$ when $n=2$. Compared with the RN-AdS black hole, the presence of parameter $\alpha$ mainly modifies the behavior of the Hawking temperature at small $x_h$. However, at large $x_h$, the curve gradually approaches that of the RN-AdS black hole. Notice that although $T$ can become negative at small $x_h$, only the positive-temperature branch is thermodynamically relevant.

Substituting Eqs.~\eqref{eq29,model mass} and \eqref{eq34,model hawking T} into Eq.~\eqref{eq28,entropy} and integrating over $x_h$ at fixed $Q$, $\Lambda$, and $\alpha$ yields the entropy
\begin{equation}
    \begin{aligned}[b]
    	S = \int_{x_0}^{x_h} \frac{1}{T(x)}\frac{\mathrm{d}M}{\mathrm{d}x}\,\mathrm{d}x &=\frac{2 \pi  x_h^2}{n} \left(-\alpha  x_h^{-n}\right)^{2/n} \Gamma \left(-\frac{2}{n},-\alpha  x_h^{-n}\right)-{S_0},\\
        S_0 &=\frac{2 \pi  x_0^2}{n} \left(-\alpha  x_0^{-n}\right){}^{2/n} \Gamma \left(-\frac{2}{n},-\alpha  x_0^{-n}\right),
        \label{eq36,model entropy}
    \end{aligned}
\end{equation}
where $\Gamma(a,z)$ denotes the upper incomplete gamma function and $S_0$ is fixed by the lower limit $x_0$ defined by $T(x_0)=0$ in Fig.~\ref{P4,model hawking T}, so that only the $T>0$ branch contributes. It can be seen that the entropy of the model deviates from the classical value $\pi{x_h}^2$. However, when $\alpha=0$, the entropy in Eq.~\eqref{eq36,model entropy} reduces to this classical value. Moreover, it is clear that the entropy is independent of both the cosmological constant $\Lambda$ and the charge $Q$. Note that although the incomplete-gamma representation is formally complex for the physical parameter range, the definite-integral entropy is strictly real, as shown in Appendix~\ref{Appendix C}.

Expanding to first order in $\alpha$ gives
\begin{equation}
    	S=\frac{1}{4} \left[A-\frac{\alpha  2^{n+1} \pi ^{n/2} A^{1-\frac{n}{2}}}{n-2}+4 \pi  x_0^2 \left(\frac{2 \alpha  x_0^{-n}}{n-2}-1\right)\right]+O[\alpha ^2],
        \label{eq37,expansion of model entropy}
\end{equation}
where $A=4 \pi{x_h}^2$ is the surface area of the black hole. For $n=2$ the expansion develops a logarithmic correction,
\begin{equation}
    	S_2 =\frac{A}{4}+ \pi  \alpha  \log \left(\frac{A}{4 \pi }\right)-2 \pi  \alpha  \log \left(x_0\right)-\pi x_0^2+O[\alpha ^2].
        \label{eq38,model entropy n=2}
\end{equation}
Such logarithmic terms are a characteristic signature of certain quantum-gravity corrections to the Bekenstein--Hawking entropy \cite{Kaul:2000kf,Carlip:2000nv,Ghosh:2006ph}. Similar logarithmic corrections have also been found in the entropy of several effective loop quantum black hole models \cite{Lin:2024flv,Ou:2025bbv}.

In AdS spacetime, the heat capacity of a black hole determines its thermodynamic stability. Specifically, the black hole is unstable when the heat capacity is negative, stable when it is positive, and undergoes a phase transition when it diverges. At fixed pressure $P$, the heat capacity of the model is
\begin{equation}
    	C=T \frac{\partial S}{\partial T} \bigg|_P=-\frac{2 \pi  x_h^2 e^{\alpha  x_h^{-n}} \left[-Q^2 x_h^n-x_h^2 \left(e^{\alpha  x_h^{-n}} \left(\alpha  n-x_h^n\right)+\Lambda  x_h^{n+2}\right)\right]}{Q^2 \left(\alpha  n-3 x_h^n\right)+\Lambda  x_h^4 \left(x_h^n+\alpha  n\right)+x_h^2
   e^{\alpha  x_h^{-n}} \left(x_h^n-\alpha  n (n+1)\right)}.
        \label{eq39,model heat capacity}
\end{equation}
Fig.~\ref{P5,model heat capacity} shows $C(x_h)$ for $n=3$ and $n=2$, truncated at $T=0$ (dashed lines) to exclude the unphysical negative-temperature regime. For fixed $Q$ and $\Lambda$, small $\alpha$ produces two divergences of $C$ with a an unstable ($C<0$) interval between them. As $\alpha$ increases, this unstable branch shrinks and eventually disappears once $\alpha>\alpha_p$, leaving a thermodynamically stable black hole.
\begin{figure}[h]
    \centering
    \hfill
    \includegraphics[width=0.49\textwidth]{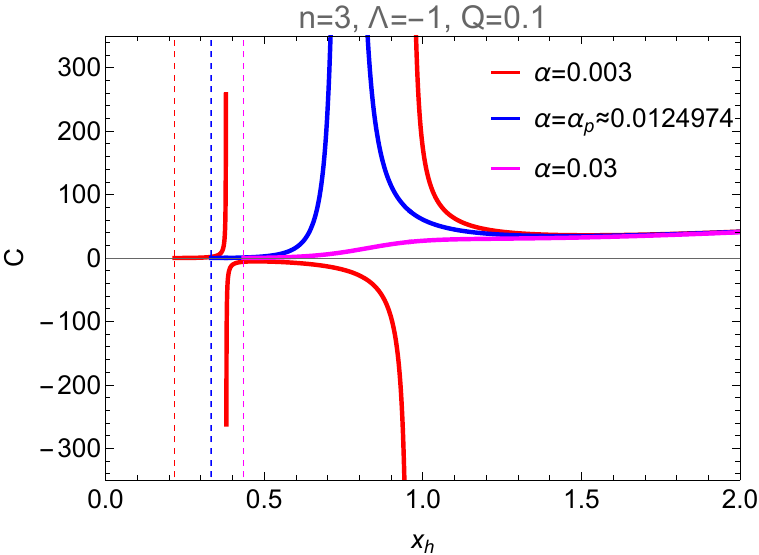}
    \hfill
    \includegraphics[width=0.49\linewidth]{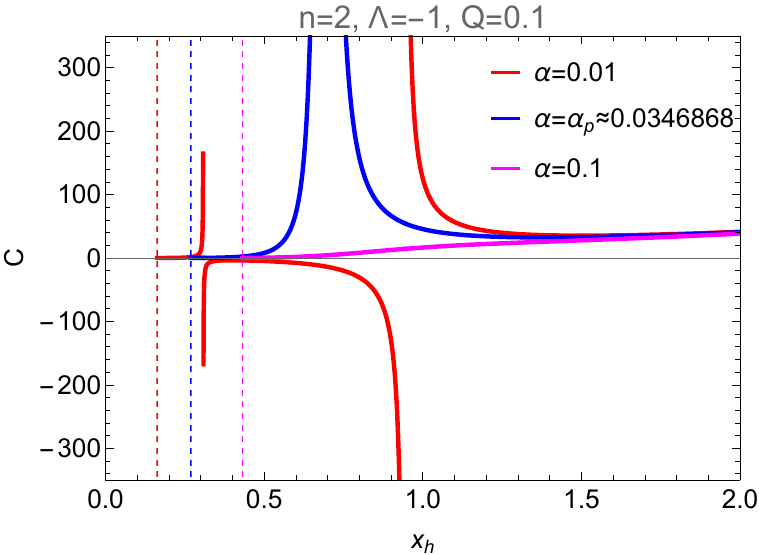}
    \caption{Plots of the heat capacity $C$ of the model as a function of $x_h$. In the left plot, the parameters are fixed at $n=3, \Lambda=-1, Q=0.1$, while $\alpha=0,0.003,0.0124974,0.03$. In the right plot, the parameters are fixed at $n=2, \Lambda=-1, Q=0.1$, while $\alpha=0,0.01,0.0346868,0.1$. The dashed line indicates the position where the Hawking temperature vanishes for the corresponding case.}
    \label{P5,model heat capacity}
\end{figure}

The free energy characterizes equilibrium among competing black-hole phases at fixed temperature and pressure. Combining Eqs.~\eqref{eq29,model mass}, \eqref{eq34,model hawking T}, and \eqref{eq36,model entropy} gives
\begin{equation}
    \begin{aligned}[b]
    	G  =& M-T S
        \\=& \frac{3 x_h^2 e^{\alpha  x_h^{-n}}-\Lambda  x_h^4+3 Q^2}{6 x_h}\\
        &+\frac{e^{-\alpha  x_h^{-n}} \left(\frac{Q^2}{x_h^3}+\Lambda  x_h\right)+\frac{\alpha  n x_h^{-n}-1}{x_h}}{4 \pi }
   \left[\frac{2 \pi  x_h^2 \left(-\alpha  x_h^{-n}\right)^{2/n} \Gamma \left(-\frac{2}{n},-\alpha  x_h^{-n}\right)}{n}-{\rm S_0}\right].
   \end{aligned}
        \label{eq40,free energy}
\end{equation}
\begin{figure}[h]
    \centering
    \hfill
    \includegraphics[width=0.49\textwidth]{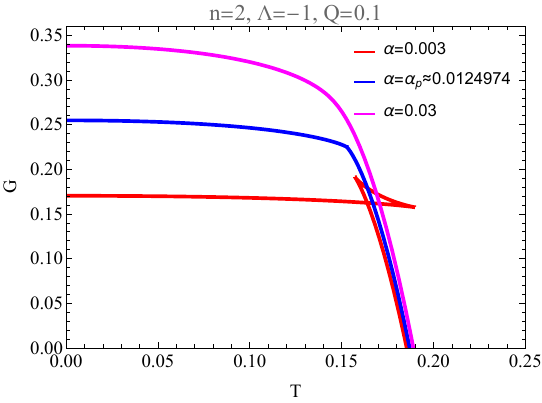}
    \hfill
    \includegraphics[width=0.48\linewidth]{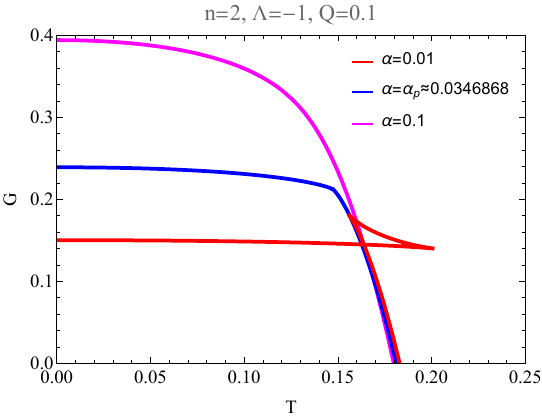}
    \caption{Plots of the free energy $G$ of the model as a function of the temperature $T$. In the left plot, the parameters are fixed at $n=3, \Lambda=-1, Q=0.1$, while $\alpha=0,0.003,0.0124974,0.03$. In the right plot, the parameters are fixed at $n=2, \Lambda=-1, Q=0.1$, while $\alpha=0,0.01,0.0346868,0.1$.}
    \label{P6,model G of T}
\end{figure}
Fig.~\ref{P6,model G of T} shows the variation of the free energy $G$ with Hawking temperature $T$ for different values of the parameter $\alpha$. When $\alpha$ is below the critical value $\alpha_p$, the plot exhibits a clear swallowtail feature, and the curve forms a self-intersecting loop, indicating that the black hole undergoes a first-order phase transition. For a given temperature, there are at most three values of $G$, corresponding to three different black hole states. When $\alpha$ exceeds the critical value, the swallowtail disappears and the curve becomes monotonic, signaling that no phase transition occurs and the black hole resides in a single stable phase.

\section{Conclusions}\label{section 4}

Within a covariant effective Hamiltonian framework, and following the reconstruction method of Ref.~\cite{Zhang:2025ccx}, we have derived the gravitational Hamiltonian constraint for the Minkowski-core regular black hole without introducing exotic matter. Setting $\alpha=0$ reduces this constraint to the classical Schwarzschild form. Coupling a spherically reduced Maxwell field and a cosmological constant then yields a charged (A)dS extension. Solving the equations of motion in the Schwarzschild and PG gauges shows that the two line elements differ only by a coordinate transformation, so the geometry is gauge independent.

Our analytical results hold for general $n$. The extended spacetime retains the defining geometric feature of the Minkowski-core model: the metric function approaches unity at small radius, and neither charge nor the cosmological constant causes the metric function to diverge as $x\to0$. Depending on the parameters, the spacetime may possess several horizons or none at all. The Kretschmann scalar vanishes as $x\to0$, indicating the absence of a curvature singularity at the center, while at large radius $\mathcal{K}$ is determined by the cosmological constant.

Our thermodynamic analysis is based on the first law of thermodynamics, rather than on an a priori assumption of the area law. The inclusion of charge, $\Lambda$, and $\alpha$ modifies the first law and yields an entropy that deviates from the Bekenstein--Hawking area law. We explicitly identified the conjugate potentials $\Phi_Q$, $\Phi_\alpha$, and the thermodynamic volume $V$, and in the limit $\alpha \to 0$, both the entropy and the first law reduce to those of the standard RN-AdS black hole, thereby restoring the area-law entropy. In particular, for $n=2$, the entropy acquires a logarithmic correction at subleading order. For a regularization parameter below a critical value the model exhibits an RN-AdS-like phase transition, which disappears once that critical value is exceeded. Minkowski-core regularization therefore preserves center regularity while qualitatively modifying the AdS phase structure relative to the RN-AdS benchmark.

It would be of interest to apply the same inverse construction to other regular black hole families, and to extend the covariant effective Hamiltonian framework beyond static configurations or to $\mathcal{R}\neq 0$.

\acknowledgments
We acknowledge the valuable discussions with Jinsong Yang. This work is supported by the Natural Science Foundation of China under Grant Nos.
12375055, 12505055, and by the Fundamental Research Funds for the Central Universities.

\appendix 
\section{Method of Construct the Hamiltonian constraint} \label{Appendix A}

This appendix reviews the reconstruction procedure of Ref.~\cite{Zhang:2025ccx}, which we use in Sec.~\ref{section 2} to obtain the effective mass $M_{\rm eff}$ and the gravitational Hamiltonian constraint $H^{\rm G}_{\rm eff}$ for a prescribed static, spherically symmetric metric---in particular, for the Minkowski-core black hole. Within this approach the target metric emerges as the unique static vacuum solution of a covariant effective theory without exotic matter.

Given that the Hamiltonian constraint $H^{\rm G}_{\rm eff}$ in Eq.~\eqref{eq9,covarianc Hamiltonian} is covariant, relevant discussions may be simplified by imposing suitable gauge conditions prior to deriving the explicit equations of motion. In the PG gauge introduced in Sec.\ref{sec:solvdynamics}, the constraint equations $H^{\rm G}_{\rm eff}=0$ and $H_x=0$ yields, with $M_{\rm eff}=M(x)$ with $M(x)$ satisfying 
\begin{equation}
	\partial_{x}M(x) + 2x\mathcal{R}(x^{2},M(x)) = 0.
    \label{eq A1,M(x)}
\end{equation}
Its solution contains an integration constant $m$, which is a Dirac observable and coincides with the mass parameter appearing in the target metric. Solving the relation $M_{\rm eff}=M(x)$ for $m$ then yields a function $m=m(M_{\rm eff},x)$, which plays a key role in reconstructing the effective Hamiltonian constraint $H^{\rm G}_{\rm eff}$ from the given metric, as described below.

As discussed in Sec.\ref{sec:solvdynamics}, preserving the PG gauge requires the lapse function $N$ and the shift vector $N^x$ to satisfy $\dot E^1=0$ and $\dot E^2=0$, yielding
\begin{equation}
    N(x) = \exp\left(2\int x(\partial_{M_{\rm eff}}R)(x)\mathrm{d}x\right),
    \label{eq A2 N(x)}
\end{equation}
\begin{equation}
    N^x(x) = -\frac{N(x)}{x}(\partial_{s_2}M_{\rm eff})(x).
    \label{eq A3 N_x(x)}
\end{equation}
Here, $\partial_{M_{\rm eff}}\mathcal{R}$ is originally a function of $(s_1,M_{\rm eff})$. In the right hand side of Eq.\eqref{eq A2 N(x)}, we need to impose the gauge condition $E^1=x^2$ together with $M_{\rm eff}=M(x)$, where $M(x)$ is determined by Eq.~\eqref{eq A1,M(x)}, thereby expressing $\partial_{M_{\rm eff}}\mathcal{R}$ solely as a function of $x$. To evaluate $N^x$, we use Eq.\eqref{eq10,covariance equations 1}, which gives
\begin{equation}
	\left( \partial_{s_2} M_{\rm eff} \right)^2 = \frac{\mu s_1}{4} \left[ (s_4)^2 + \mathcal{Z} \right],
    \label{eq A4,covariance equations simplifies}
\end{equation}
where $\mathcal{Z}=\mathcal{Z}(s_1,M_{\rm eff})$ is an arbitrary function characterizing the effective Hamiltonian. Substituting this relation into Eq.\eqref{eq A3 N_x(x)} and imposing again the gauge condition $E^1=x^2$ together with $M_{\rm eff}=M(x)$, we obtain $N^x$ as a function of $x$.

By transforming from PG to Schwarzschild coordinates via ${\rm d}t_s = {\rm d}t_p + \frac{4 N^x}{\mu N^2 \mathcal{Z}} {\rm d}x$, the general metric \eqref{eq8,general metric} is matched to a prescribed static metric in Schwarzschild coordinates $(t_s,x,\theta,\phi)$,
\begin{equation}
    \begin{aligned}[b]
        \mathrm{d}s^2 &= \frac{1}{4} N^2\mathcal{Z}\mathrm{d}t_s^2 -\frac{4}{\mu\mathcal{Z}}\mathrm{d}x^2 +x^2\mathrm{d}\Omega^2 \\
    &= -g_0(x;m)\mathrm{d}t_s^2 +g_1(x;m)^{-1}\mathrm{d}x^2 +x^2\mathrm{d}\Omega^2.
    \end{aligned}
    \label{eq A5,static metric }
\end{equation}
Comparing the two forms in Eq.~\eqref{eq A5,static metric } and substituting into Eq.~\eqref{eq A4,covariance equations simplifies} with $\sqrt{s_1}=x$ yields the integral equation for $M_{\rm eff}$:
\begin{equation}
    \begin{aligned}[b]
        &\pm \frac{\sqrt{s_1}}{2} s_2 + \Xi (s_1,s_4)\\ =& \int^{M_{\mathrm{eff}}}\mathrm{d}\xi \Bigg\{\frac{N(\sqrt{s_1})^2g_{1}(\sqrt{s_1};m(\xi,\sqrt{s_1}))(s_4)^2}{g_{0}(\sqrt{s_1};m(\xi,\sqrt{s_1}))} -4g_{1}(\sqrt{s_1};m(\xi,\sqrt{s_1}))\Bigg\}^{-1 / 2},
    \end{aligned}
    \label{eq A6,integral equation of M}
\end{equation}
where $\Xi$ is an integration constant. Inserting the known metric functions $g_0$ and $g_1$ into \eqref{eq A6,integral equation of M} determines $M_{\rm eff}$, and substitution into \eqref{eq9,covarianc Hamiltonian} then yields $H^{\rm G}_{\rm eff}$.

As a check, for the Schwarzschild metric $g_0=g_1=1-2m/x$ with $\mathcal{R}=0$ and $\Xi=0$, Eq.~\eqref{eq A6,integral equation of M} gives
\begin{equation}
	M_{\rm eff}=\frac{1}{8} \sqrt{s_1} \left[4 (s_2)^2-(s_4)^2+4\right].
    \label{eq A7,Meff of sch}
\end{equation}
For the Minkowski-core model discussed in Sec.~\ref{section 2}, we adopt the same choice of integration constants. Because of residual freedom in these constants, the constructed effective Hamiltonian constraint $H^{\rm G}_{\rm eff}$ is not unique. However, for a given Hamiltonian constraint, the metric derived from it is unique. Finally, a practical limitation of this method is that Eq.~\eqref{eq A6,integral equation of M} does not always admit an explicit closed-form solution for $M_{\rm eff}$.

\section{Equations of motion} \label{Appendix B}

In this appendix, we derive the equations of motion associated with the total Hamiltonian constraint $H^{\rm T}_{\rm eff}$ introduced in Sec.~\ref{section 2B}, which includes couplings to a Maxwell field and a cosmological constant. The vacuum Minkowski-core constraint and the corresponding metric are recovered when $Q=\Lambda=0$. The phase-space variables $(E^1,E^2,K_1,K_2)$, together with the lapse function $N$ and shift vector $N_x$, satisfy four evolution equations and the diffeomorphism and Hamiltonian constraints. A gauge choice then determines the line element. We first derive the equations of motion associated with the total Hamiltonian constraint in the Schwarzschild gauge and obtain the corresponding metric \eqref{eq21,new metric}. We then carry out the corresponding analysis in the PG gauge to obtain \eqref{eq23,new metric PG}. Throughout, $x$ denotes the radial coordinate and a prime indicates a derivative with respect to $x$.

In the Schwarzschild gauge $\partial_{s_2}M_{\rm eff}=0$, $E^1(x)=x^2$, $M_{\rm eff}$ can be rewritten as $K_2(x) \, x\, e^{\alpha (-x^{-n})}=0$. For this equality to hold identically, we must set $K_2(x)=0$. Then the diffeomorphism constraint $H^{\rm T}_x=0$ implies $K_1(x)=0$, and furthermore the Hamiltonian constraint $H^{\rm T}_{\rm eff}=0$ becomes a differential equation for $E^2(x)$:
\begin{equation}
    \begin{aligned}[b]
0=&2 x^4  {E^2}'(x) \ e^{\alpha  x^{-n}}+E^2(x)^3 \ x^{-n-1} \left[-Q^2 x^n-\Lambda  x^{n+4}-x^2 e^{\alpha  x^{-n}} \left(\alpha  n-x^n\right)\right]\\
&-E^2(x) \ x^{3-n}
   e^{\alpha  x^{-n}} \left(3 x^n-\alpha  n\right).
   \end{aligned}
    \label{eq B1,E2 eq schw}
\end{equation}
Solving this equation yields
\begin{equation}
E^2(x)=-\frac{x^{7/2} e^{\frac{\alpha  x^{-n}}{2}}}{\sqrt{x^5 e^{\alpha  x^{-n}}+Q^2 x^3-\frac{\Lambda  x^7}{3}+c_1 x^4}},
    \label{eq B2,E2 sol schw}
\end{equation}
where $c_1$ is an integration constant. Substituting the above result into $\dot E^1(x)=0$ and requiring the equality to hold identically gives $N^x(x)=0$. $\dot K_2(x)=0$ reduces to a differential equation for $N(x)$:
\begin{equation}
    \begin{aligned}[b]
0=&2 N'(x) \ \left[3 Q^2+x \left(3 x e^{\alpha  x^{-n}}-\Lambda  x^3+3 c_1\right)\right]\\
&+x^{-n-1} N(x) \ \left[Q^2 \left(6 x^n-3 \alpha  n\right)+x \left(-3 \alpha 
   c_1 n+2 \Lambda  x^{n+3}+3 c_1 x^n+\alpha  \Lambda  n x^3\right)\right].
    \end{aligned}
    \label{eq B3,N eq schw}
\end{equation}
Solving it gives 
\begin{equation}
N(x)=\frac{c_2 e^{-\frac{1}{2} \alpha  x^{-n}} \sqrt{3 Q^2+x \left(3 x e^{\alpha  x^{-n}}-\Lambda  x^3+3 c_1\right)}}{x},
    \label{eq B4,N sol schw}
\end{equation}
with $c_2$ another integration constant. In order for the model to recover the classical case, the integration constants are chosen as $c_1=-2M$ and $c_2=\frac{1}{\sqrt{3}}$. Substituting $E^1=x^2$, $E^2$ from \eqref{eq B2,E2 sol schw}, $N^x=0$, and $N$ from \eqref{eq B4,N sol schw} into the general metric \eqref{eq8,general metric} then yields the Schwarzschild-gauge line element \eqref{eq21,new metric} of Sec.~\ref{section 2B}.

We next solve the same constraints in the PG gauge $E^1(x)=x^2$ and $E^2(x)=x$. Imposing $H^{\rm T}_x=0$ and $H^{\rm T}_{\rm eff}=0$ gives
\begin{equation}
    K_2(x)=\frac{e^{\frac{\alpha  x^{-n}}{2}} \sqrt{-Q^2+\frac{\Lambda  x^4}{3}+C_1 x}}{x},
    \label{eq B5,PG K2}
\end{equation}
\begin{equation}
    K_1(x)=\frac{x^{-n-2} e^{\frac{\alpha  x^{-n}}{2}} \left[\alpha  n \left(3 Q^2-\Lambda  x^4-3 C_1 x\right)+x^n \left(6 Q^2+2 \Lambda  x^4-3 C_1 x\right)\right]}{2 \sqrt{-9 Q^2+3 \Lambda  x^4+9 C_1 x}},
    \label{eq B6,PG K1}
\end{equation}
where $C_1$ is an integration constant. Owing to the choice of PG gauge, $E^1(x)$ and $E^2(x)$ are both independent of $t_p$. From $\{E^1(x), H^{\rm T}_{\rm eff}[N]+H^{\rm T}_x[N^x]\} = 0$ and $\{E^2(x), H^{\rm T}_{\rm eff}[N]+H^{\rm T}_x[N^x]\} = 0$, we have
\begin{equation}
    N^x_p(x)=-\frac{N e^{-\frac{1}{2} \alpha  x^{-n}} \sqrt{-Q^2+\frac{\Lambda  x^4}{3}+C_1 x}}{x},
   \label{eq B7,PG Nx}
\end{equation}
\begin{equation}
    N(x)=C_2,
    \label{eq B8,PG N}
\end{equation}
where $C_2$ is another integration constant. Then we can determine the metric function \eqref{eq23,new metric PG} in PG coordinates. To ensure that the model reduces to the case of GR when the constant $\alpha$ tends to zero, the integration constants are chosen as $C_2=1$ and $C_1=-2m$, respectively. Inserting these PG solutions into \eqref{eq8,general metric} yields the PG line element \eqref{eq23,new metric PG} quoted in the main text.

\section{Kretschmann scalar and reality of the entropy} \label{Appendix C}

This appendix collects two technical ingredients used in the main text: the complete expression for the Kretschmann scalar underlying the regularity analysis of Sec.~\ref{section 2C}, and a brief argument that the entropy of Sec.~\ref{section 3} is real despite the appearance of an incomplete gamma function.

As stated in Sec.~\ref{section 2C}, the Kretschmann scalar of the extended model can be written as \eqref{eq25,model K scalar}. Expanding the polynomial $\mathcal{P}$ yields the explicit formula
\begin{equation}
    \begin{aligned}[b]
        \mathcal{K}=&\frac{1}{9} x^{-4 (n+2)} e^{-2 \alpha  x^{-n}} \bigg\{36 M^2 x^2 \bigg[\alpha ^4 n^4-2 \alpha ^3 n^3 (n+3) x^n \\ &+\alpha ^2 n^2
   (n^2+6 n+17) x^{2 n}-4 \alpha  n (n+5) x^{3 n}+12 x^{4 n}\bigg]\\
   &+12 \alpha  \Lambda  M n x^5 \bigg[\alpha ^3 n^3-2
   \alpha ^2 n^3 x^n+\alpha  n (n^2-1) x^{2 n}-4 (n-1) x^{3 n}\bigg]\\
   &-6 Q^2 \bigg[6 M x \bigg(\alpha ^4 n^4-2 \alpha ^3 n^3
   (n+4) x^n+\alpha ^2 n^2 (n^2+8 n+27) x^{2 n}\\
   &-8 \alpha  n (n+5) x^{3 n}+24 x^{4 n}\bigg)+\alpha  \Lambda  n x^4
   \bigg(\alpha ^3 n^3+\alpha  n (n^2+2 n-3) x^{2 n}\\
   &-2 \alpha ^2 n^2 (n+1) x^n-8 (n-1) x^{3 n}\bigg)\bigg]+9 Q^4
   \bigg[\alpha ^4 n^4-2 \alpha ^3 n^3 (n+5) x^n\\
   &+\alpha ^2 n^2 (n^2+10 n+41) x^{2 n}-4 \alpha  n (3 n+19) x^{3 n}+56 x^{4
   n}\bigg]\\
   &+\Lambda ^2 x^8 \bigg[\alpha ^4 n^4-2 \alpha ^3 (n-3) n^3 x^n+\alpha ^2 n^2 (n^2-6 n+17) x^{2 n}\\
   &-4 \alpha (n-7) n x^{3 n}+24 x^{4 n}\bigg]\bigg\}.
    \end{aligned}
    \label{eq C1,A}
\end{equation}
The asymptotic limits quoted in \eqref{eq26,K scalar limit} follow directly from this expression.

The entropy of the extended model obtained from the indefinite integral \eqref{eq28,entropy} contains the upper incomplete gamma function $\Gamma(-\frac{2}{n},-\alpha  x_h^{-n})$. For the physical range $\alpha>0$, $n>0$, and $x_h>0$, this gamma function is in general complex-valued. However, differentiating \eqref{eq36,model entropy} with respect to $x_h$ yields $2\pi x_h\,e^{\alpha/x_h^{\,n}}$, 
which is purely real for all $x_h>0$. Consequently,
\begin{equation}
    \frac{{\rm d}}{{\rm d}x_h}\operatorname{Im}\bigl[S(x_h)\bigr]=\operatorname{Im}\!\left[\frac{{\rm d}S(x_h)}{{\rm d}x_h}\right]=0,
\end{equation}
meaning that $\operatorname{Im}[S(x_h)]$ is a constant independent of $x_h$ on the positive real axis.

In the definite integral from $x_0$ to $x_h$ that defines the physical entropy, this constant imaginary part cancels between the two endpoints. The entropy employed in Sec.~\ref{section 3} is therefore strictly real.

\bibliographystyle{utphys}
\bibliography{Ref}
\end{document}